\documentclass[prd,twocolumn,amssymb,amsmath]{revtex4}
\usepackage[applemac]{inputenc}
\usepackage{graphicx}
\usepackage{txfonts}

\usepackage{color}

\begin{document}
\newcommand{\bx}{{\bf x}}
\newcommand{\bn}{{\bf n}}
\newcommand{\bk}{{\bf k}}
\newcommand{\dd}{{\rm d}}
\newcommand{\dslash}{D\!\!\!\!/}
\def\ga{\mathrel{\raise.3ex\hbox{$>$\kern-.75em\lower1ex\hbox{$\sim$}}}}
\def\la{\mathrel{\raise.3ex\hbox{$<$\kern-.75em\lower1ex\hbox{$\sim$}}}}
\def\beq{\begin{equation}}
\def\eeq{\end{equation}}

\vskip-2cm
\title{\textcolor{black}{ Axions, Time Varying CP Violation, and Baryogenesis}}

\author{Lawrence M. Krauss}
\affiliation{
   The Origins Project Foundation, Phoenix AZ 85020}
\email{lawrence@originsprojectfoundation.org}

\vspace*{1cm}

\begin{abstract} 

We derive two features of axion cosmology that may have cosmological implications, whether or not axions are dark matter:  For the full range of allowed axion masses, the evolution of a cosmic axion background allows large CP violation until temperatures as low as $\sim$ 2 GeV, and once the axion field begins to oscillate, the cosmological axion field's relaxation to its ground state can briefly provide a new departure from thermal equilibrium, via time-varying CP violation.  During both of these periods, the Strong CP violating parameter $\bar\theta$ can be as large as O(1).
\end{abstract}

\date{\today}

\maketitle

%%%%%%%%%%%%%%%%%%%%%%%
%%%%%%%%%%%%%%%%%%%%%%%
%%%%%%%%%%%%%%%%%%%%%%%
%%%%%%%%%%%%%%%%%%%%%%%
%%%%%%%%%%%%%%%%%%%%%%%
%%%%%%%%%%%%%%%%%%%%%%%
%%%%%%%%%%%%%%%%%%%%%%%
%%%%%%%%%%%%%%%%%%%%%%%
%%%%%%%%%%%%%%%%%%%%%%%
%%%%%%%%%%%%%%%%%%%%%%%
%%%%%%%%%%%%%%%%%%%%%%%

%%%%%%%%%%%%%%%%%%%%%%%
%%%%%%%%%%%%%%%%%%%%%%%
%%%%%%%%%%%%%%%%%%%%%%%
%%%%%%%%%%%%%%%%%%%%%%%
%%%%%%%%%%%%%%%%%%%%%%%
%%%%%%%%%%%%%%%%%%%%%%%
%%%%%%%%%%%%%%%%%%%%%%%
%%%%%%%%%%%%%%%%%%%%%%%
%%%%%%%%%%%%%%%%%%%%%%%
%%%%%%%%%%%%%%%%%%%%%%%
%%%%%%%%%%%%%%%%%%%%%%%

\section{Introduction} 
\label{introduction}

Sakharov first demonstrated \cite{sakharov} that the observed baryon number of the universe may be dynamically generated as the Universe evolves if three conditions are present: (1) baryon number violating interactions, (2) departure from thermal equilibrium, and (3) CP violation.  Since his original proposal, numerous scenarios have been proposed in which all three conditions exist at early cosmic times.  The standard model of particle physics in fact incorporates all three possibilities but in general cannot generate the observed baryon number today because of two factors, small CP violation, and small departures from thermal equilibrium. For this reason it has to be supplemented by new physics in order to produce a phenomenologically acceptable value of $n_B$ today. 

Axions \cite{pq} were proposed to explain the absence of observed CP violation in the Strong Interaction, by providing a dynamical mechanism for $\bar{\theta}$---the effective Strong Interaction CP violating parameter arising from a combination of a non-perturbative topological term $ {\theta \over 32\pi^2} \text{Tr} G_{\mu\nu}\tilde{G}^{\mu\nu}$ and a possible imaginary phase $\theta'$ that may appear in quark mass matrix---to relax to zero by the present time.  Specifically $\bar{\theta}$ becomes proportional to the axion field, where the axion is a pseudo-Goldstone boson associated with a new spontaneously broken axial symmetry, the minimum of whose potential lies at $a=0$ 

This same dynamics, however, implies that cosmological axions inherently incorporate two of Sakharov's three conditions in the early universe, and so may be relevant for baryogenesis or have other cosmological implications: possible large CP violation until relatively low temperatures, and a departure from equilibrium involving time-varying CP violation as $\bar{\theta}$ relaxes to zero.

\section{Early Universe Axion Dynamics} 

Axions arise from the breaking of a new global axial symmetry at high energy, which, through instanton effects associated with  the axial anomaly, turns the CP violating $\bar{\theta}$ into a dynamical quantity. The axion is the angular mode of a complex scalar field whose non-zero radial expectation value breaks the axial PQ symmetry at some scale $\sim f_a$.  The angular potential is initially flat and the value of the axion field is undetermined, with the appropriately normalized dimensionless field taking some value between -$\pi$ and $\pi$.  If some initially small region in which the axion field is uniform inflates, a coherent uniform background axion will result today.  In non-inflationary models, quantum fluctuations in the axion field are unconstrained, and while the value can therefore spatially vary, the RMS spatially averaged value of the field will be non-zero.  

At late times, due to chiral symmetry breaking when the temperature of the Universe is $\sim$ O(1 GeV), the PQ symmetry is explicitly broken, turning the axion into a Pseudo-Goldstone boson with mass calculable in chiral perturbation theory to be \cite{ww}
\begin{equation}
m_a \approx  6 \times 10^{-6} \ \  \text{eV} \ \ ({10^{12} \  \text{GeV} \over f_a })
\label{eq:111}
\end{equation}
where $f_a$ is, by analogy to pion physics, the decay constant of the axion.

In the Peccei-Quinn scenario, $\bar{\theta (t)} \approx {a(t) \over f_a}$  so that $\bar{\theta}$ dynamically relaxes to zero as the cosmological axion field $a(t)$ relaxes to the minimum of its potential.  

Note first that since initially $a(t)$ is unconstrained, $\bar{\theta}$ can be as large as order unity initially, allowing for large CP violating effects. Moreover, as we now illustrate, these effects will in general persist until temperatures well below the electroweak phase transition.

A classical axion field with initial non-zero value $a_0$ will generically oscillate about 0 with characteristic frequency $\sim  m_a$.  Cosmologically, an axion field will evolve via the classical equation of motion 

\begin{equation}
\ddot{a} + 3H\dot{a} +m_a^2 a = 0
\end{equation}

This leads to the following canonical behavior of an axion field \cite{dm,marsh,LMKtasi}.  When $H >> m$, $\dot{a}  \approx 0$, while when $m >> H$, $a(t) \sim a_0$ cos ($mt)$.  This can be understood heuristically as follows.  When $H >>m$ the age of the universe is much smaller than the natural period of oscillation of the axion field, so it remains constant.  When $H << m$ the expansion of the universe can be ignored, and the axion field oscillates at its natural frequency.   To understand what happens when $H \sim m$, we can consider more generally the equations for energy density and pressure associated with the axion field: 
\begin{equation}
\rho_a = {1 \over 2} \dot{a}^2 + {1 \over 2} m_a^2 a^2
\end{equation} 
\begin{equation}
P_a = {1 \over 2} \dot{a}^2 - {1 \over 2} m_a^2 a^2
 \end{equation} 

At early times, when $\dot{a}=0$, the (constant) axion field behaves like a cosmological term, with $P=-\rho$.  As a result, its energy density doesn't redshift, allowing it to become, in principle, cosmologically significant at late times, so that axions can be dark matter candidates.  However, once the axion field begins to oscillate with period $m$, the pressure approaches zero, meaning that the classical axion field has the equation of state of a pressureless, non-relativistic gas.    We know that the energy density of such a gas redshifts inversely as the cube of the scale factor $R$.  Since the energy density of the axion field is $\sim m^2a^2$, this implies that ${a} \sim R^{-3/2}$.  Assuming that the expansion of the universe is radiation-dominated that this time, then $R \sim t^{1/2}$, implying that  $a \sim t^{-3/4}$.

Note also that because the axion field remains constant until $H \sim m$, the energy density stored in the axion field also remains roughly constant until $H \sim m$.  Thus the remnant axion energy density today turns out to be inversely proportional to the axion mass, ignoring for the moment any temperature dependence of the axion mass.  For an initial value ${a_0 \over f_a}  \sim 1$, the cosmological energy density in axions would exceed the inferred dark matter density today if $f_a > 4 \times 10^{12}$ GeV \cite{pq}.    At the same time, astrophysical constraints on cosmic axions, particularly limits on white dwarf cooling \cite{marsh,raffelt,PDG}, imply $ f_a > O(10^{10})$ GeV, seemingly squeezing the allowed range of axion dark matter.   

However, accepting the fact that Inflation generically leads to a multiverse, it has been recognized that the actual value of ${a_0\over f_a}$ in our universe could be smaller than O(1).   Assuming, for example, that the Peccei-Quinn scale and the Inflation scale were both of O($10^{16}$) GeV, this would imply an axion mass of O($10^{-10}$)  eV, and in order not to overclose the universe today, ${a_0 \over f_a}  \le$ O $(10^{-4})$ . 

\section{Large CP Violation for $T > $O(1) GeV } 

One might expect that the time when the axion field begins to relax to its minimum, and thus CP violation would quickly drop to zero, would vary inversely with axion mass.  However, in QCD, chiral symmetry breaking occurs at a temperature close to the confinement scale $\Lambda_{QCD} \sim $ O(1) GeV.  Instanton effects that give the axion a non-zero mass typically vanish at temperatures well above this scale.   Using a dilute-instanton-gas approximation one finds \cite{dm}

\begin{equation}
m_a(T) \sim m_a(0) T^{-4}[ln(T)]^3 \ \ \text{for} \ \ T >  \Lambda. 
\end{equation}  

During the radiation dominated phase at $T \sim \Lambda $, $ t \sim {T^2 \over M_{pl}}$, where $M_{pl}$ is the Planck mass.  Thus, if the  axion field begin to relax when $t_i \sim m_a(T) $, then from eq. (5) we find that for axions whose natural oscillation time is less than the time when the universe undergoes its chiral phase transition, the temperature at which axion oscillations begin scales as

\begin{equation}
T \sim f_a^{-1/6}   
\end{equation}
rather than as $f_a^{-1/2}$.

In standard cosmology, the age of the universe when chiral symmetry breaks, when $T \sim 200$ MeV, is approximately $5 \times 10^{-5}$ s.  This corresponds to an oscillation period for an axion with a mass $\sim 10^{-10}$ eV and $f_a \sim 10^{16}$ GeV.     

If we consider that astrophysics currently puts a lower bound on $f_a$ of $\sim 10^{10}$ GeV, this means that for viable axion models the axion field does not begin to relax from its initial post-Inflationary value until at most a temperature $ T \sim 2$ GeV.  This means that $\bar{\theta}$, and hence Strong CP violation, remains relatively large, namely greater than $\sim 10^{-7}$ for $f_a < M_{Pl}$ and as large as O(1) for $f_a < 10^{12}$ GeV, until well below even the electroweak phase transition.  

This upper bound on the relaxation temperature at which $\bar\theta$ begins to decrease means that even low energy baryogeneis \cite{largeCP,LEB} scenarios will benefit from relaxed constraints on new sources of CP violation.  The extent to which large Strong Interaction CP violation would affect different scenarios is of course model-dependent.  But this result suggests that if a cosmological axion field exists, even if it does not comprise the inferred dark matter dominating galaxies today, other new sources of CP violation beyond the standard model may not be required for successful baryogenesis.     
 
\section{Time Varying CP Violation when  $H \sim m_a$} 

An interesting new phenomenon occurs for axion models in the allowed range, $ f_a > 10^{10}$ GeV, corresponding to a relaxation temperature below 2 GeV.  For these masses, the axion field will begin to oscillate with a period comparable to the age of the Universe at that time.  For the range $10^{10} < f_a < 10^{16}$ GeV the axion mass will be temperature dependent between 2 GeV and the chiral symmetry breaking scale as we have seen.  For lighter axions, which begin to relax below the chiral symmetry breaking scale, their mass will already be fixed by the time oscillations start.

For simplicity, we consider here an axion field with $f_a \sim 10^{16}$, which begins to relax just after the chiral symmetry breaking transition is complete, although the effect we discuss will be qualitatively similar, if slightly more complex to compute, for other masses.  This scale is also of interest as it corresponds to a possible scale of Grand Unification in various supersymmetric theories.  

As the axion field begins to relax, via damped oscillations, to its preferred ground state on a timescale comparable to the age of the Universe (and the inverse Hubble constant), this represents a non-equilibrium phenomenon, much like the late-time out-of-equilibrium decay of a long-lived particle.  In this case, however, the time-varying order parameter describing the out-of-equilibrium transition is the $\bar\theta$ parameter of QCD.   If the field were merely oscillating and not redshifting, then oscillations in $\bar\theta$ would average to zero.  However, the damping of oscillations implies that this will not be the case.  

For processes that take place on a timescale commensurate with $H^{-1}$ we can visualize the oscillations in the axion field, and hence in $\bar\theta$, as short non-equilibrium pulses of non-zero CP violation.   As shown in Section II, the amplitude of the axion field's oscillations, and thus $\bar\theta$, falls as $t^{-3/4}$ once the field begins to oscillate.  We can estimate the average magnitude of the time-varying CP violation contribution to ongoing processes during this period by adding up the the successive oscillating non-zero CP pulses. The sum of a series of oscillating pulses falling off as $t^{-3/4}$ is given by the Dirichlet $\eta$ function with argument $3/4$:

\begin{equation}
\eta(3/4) = \sum_{n=1}^{+\infty} {-1^{n-1} \over t^{3/4}} \approx 0.651
\end{equation}
                                                                                                                                                                                                                                                                                                                                                                                                                                                                                                                                                                                                                                                                                                                                                                                                                                                                                                                                                                                                                                                       
Thus, the effective magnitude of CP violation during this non-equilibrium relaxation of the axion field is compared to the initial value before the onset of oscillation.  

\section{Conclusion}

Independent of their mass, axions allow both significant CP violation effects in the early universe and also allow for a brief out-of-equilibrium relaxation period involving time-varying Strong CP violation.  The former effect could significantly impact baryogenesis in the early universe, including low-energy baryogenesis scenarios.   Whether the latter period of time-varying CP violation is merely an interesting curiosity, or may have some residual impact on cosmological or astrophysical parameters may be worthy of further study.

%%%%%%%%%%%%%%%%%%%%%%%
%%%%%%%%%%%%%%%%%%%%%%%
%%%%%%%%%%%%%%%%%%%%%%%
%%%%%%%%%%%%%%%%%%%%%%%
%%%%%%%%%%%%%%%%%%%%%%%
%%%%%%%%%%%%%%%%%%%%%%%
%%%%%%%%%%%%%%%%%%%%%%%
%%%%%%%%%%%%%%%%%%%%%%%
%%%%%%%%%%%%%%%%%%%%%%%
%%%%%%%%%%%%%%%%%%%%%%%
%%%%%%%%%%%%%%%%%%%%%%%

%%%%%%%%%%%%%%%%%%%%%%%
%%%%%%%%%%%%%%%%%%%%%%%
%%%%%%%%%%%%%%%%%%%%%%%
%%%%%%%%%%%%%%%%%%%%%%%
%%%%%%%%%%%%%%%%%%%%%%%
%%%%%%%%%%%%%%%%%%%%%%%
%%%%%%%%%%%%%%%%%%%%%%%
%%%%%%%%%%%%%%%%%%%%%%%
%%%%%%%%%%%%%%%%%%%%%%%
%%%%%%%%%%%%%%%%%%%%%%%
%%%%%%%%%%%%%%%%%%%%%%%

\end{document}